\documentclass[%
 reprint,
 amsmath,amssymb,
 aps,soul,
pre,
]{revtex4-2}

\usepackage{graphicx}
\usepackage{dcolumn}
\usepackage{bm}
\usepackage{xcolor}

\renewcommand{\selectlanguage}[1]{}
\usepackage{lipsum}
\makeatletter
\newcommand*{\balancecolsandclearpage}{%
  \close@column@grid
  \cleardoublepage
  \twocolumngrid
}
\begin{document}

\preprint{APS/123-QED}

\title{Kinetic Turing Instability and Emergent \\ Spectral Scaling in Chiral Active Turbulence}



\author{Magnus F Ivarsen}
\email{Contact: magnus.fagernes@gmail.com}
\altaffiliation[Also at ]{
The European Space Agency Centre for Earth Observation, Frascati, Italy}
\affiliation{Department of Physics and Engineering Physics, University of Saskatchewan, Saskatoon, Canada}%

\begin{abstract}
The spontaneous emergence of coherent structures from chaotic backgrounds is a hallmark of active biological swarms. We investigate this self-organization by simulating an ensemble of polar chiral active agents that couple locally via a Kuramoto interaction. We demonstrate that the system's transition from chaos to active turbulence is characterized by quantized loop phase currents and coherent clustering, and that this transition is strictly governed by a kinetic Turing instability. By deriving the continuum kinetic theory for the model, we identify that the competition between local phase-locking and active agent motility selects a critical structural wavenumber. The instability then drives the system into a state of developed, active turbulence that exhibits stable, robust power-laws in spectral density, suggestive of universality and consistent with observations from a broad range of turbulent phenomena. Our results bridge the gap between discrete chimera states and continuous fluid turbulence, suggesting that the statistical scaling laws of active turbulence can arise from fundamental kinetic instability criteria.
\end{abstract}


\maketitle


\section{Introduction}

The emergence of macroscopic order from chaotic microscopic constituents defines complex systems \cite{laughlin_different_2005}, from turbulent geophysical \cite{wojcik_searching_2025}, and astrophysical \cite{marov_self-organization_2013} structuring, to superconductivity \cite{keimer_high_2014,bussmann-holder_high-temperature_2020}, soft (active) matter dynamics \cite{wensink_meso-scale_2012,aranson_bacterial_2022}, and active biological swarms \cite{toner_flocks_1998,wensink_emergent_2012,giomi_defect_2013}. Common for these phenomena are complex many-body interactions, described as non-linear mode coupling in turbulence physics \cite{ochiSpectralAnalysis1998}, wherein the ordered complexity of the greater system is subject to statistical mechanics \cite{fromm_emergence_2004,marov_self-organization_2013,zheng_introduction_2021}. 

In this letter, we investigate the emergence of order in active matter turbulence, by modeling active agents that synchronize locally. While local synchronization is a known driver of coherent structuring in active matter turbulence \cite{acebron_kuramoto_2005,okeeffe_oscillators_2017,hong_active_2018,okeeffe_oscillators_2017}, the precise kinetic mechanisms that spontaneously select specific length scales in continuous, active media remain elusive.

Recent advances in network science describe pattern formation, such as chimera states, through topological constraints on fixed network lattices \cite{luo_turing_2023}. However, network models neglect the motion (transport) inherent to active matter, where agent motility functions as a kinetic inhibitor competing against local phase-locking activation \cite{alert_active_2022,marmol_colloquium_2024,maitra_activity_2025}.

We bridge this gap by simulating polar chiral active agents subject to localized Kuramoto interactions. We demonstrate that the resulting active turbulence arises strictly from a kinetic Turing instability \cite{turing_chemical_1990}, where the competition between active motility and phase synchronization selects a critical wavenumber, generating stable and possibly universal power-law scaling in spectral density that closely aligns with routine turbulence observations.

\section{\label{sec:model}The model}
\vspace{-8pt}




Motivated by an astrophysical model of active, excitable fluid systems of interacting elements capable of self-organization  \cite{marov_self-organization_2013}, we modeled chiral active matter as an ensemble of $N$ self-propelled agents inside a 2-dimensional (2D) bounded box. The agents' orientation is decided entirely by their internal phase $\phi_i\in [0, 2\pi)$,
\begin{equation} \label{eq:velocity}
    \mathbf{v}_i(t) = v_0  (\cos\phi_i(t), \; \sin\phi_i(t)),
\end{equation}
with $v_0$ being the active agents' constant swim speed, and with the evolution of the $i^\text{th}$ agent decided chiefly by its intrinsic \textit{frustration} $\omega_i$, a synchronization rule, and a Gaussian noise term $\eta_i$,
\begin{equation} \label{eq:kuramoto}
    \dot{\phi}_i = \underbrace{\omega_i}_{\text{Driver}} + \underbrace{a_0 R(\mathbf{x}_i,t) \sin(\Psi(\mathbf{x}_i,t) - \phi_i)}_{\text{Synchronization force}} + \underbrace{\eta_i(t)}_{\text{Noise}},
\end{equation}
where we identify the synchronization force as a locally mediated Kuramoto-Sakaguchi-coupling \cite{acebron_kuramoto_2005,de_smet_partial_2007}. $\Psi(\mathbf{x}, t)$ is the local mean phase, determined by the complex local order parameter field,
\begin{multline} \label{eq:kernel}
    Z(\mathbf{x}, t) = R(\mathbf{x}, t) e^{i\Psi(\mathbf{x}, t)} = \\ = \int_{\mathcal{D}} G(|\mathbf{x} - \mathbf{x}'|) \left[ \sum_{j=1}^N \delta(\mathbf{x}' - \mathbf{x}_j(t)) e^{i\phi_j(t)} \right] d\mathbf{x}',
\end{multline}
and we note that $R(\mathbf{x}_i,t)$ in Eq.~(\ref{eq:kuramoto}) is the amplitude of the complex local mean field $Z$, which is  defined as the convolution of the microscopic agent distribution with a finite-range interaction kernel $G(|\mathbf{x}-\mathbf{x}'|)$. $G$ represents the Green's function of the agents' interaction (modeled as a Gaussian with kernel size $\sigma$), mediating chemical interactions or electric fields, which are taken to govern the dense active matter interactions.


\begin{figure}
    \centering
    \includegraphics[width=0.5\textwidth]{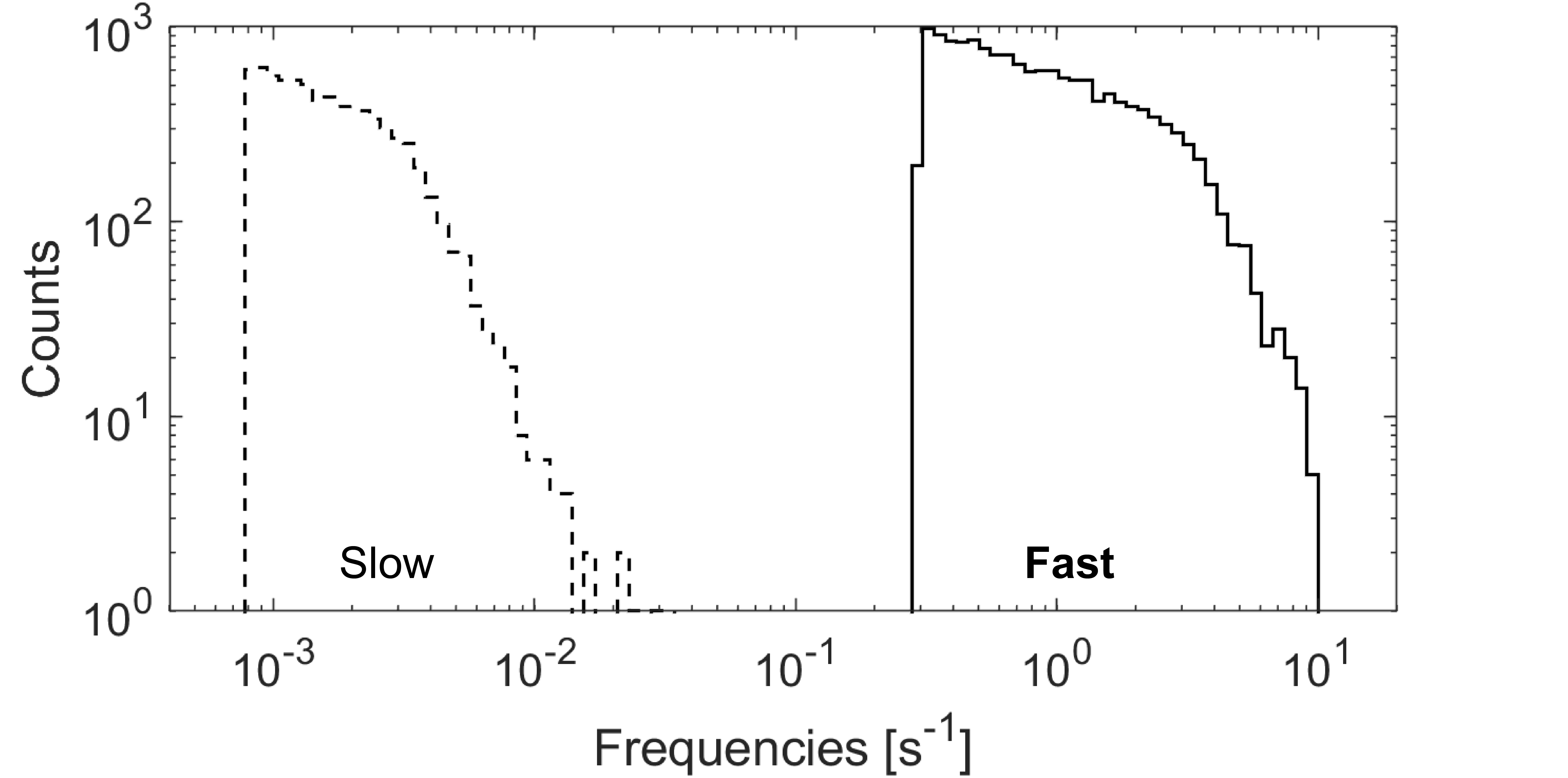}
    \caption{Slow (dashed line) and fast (solid line) distributions in intrinsic frustration $\omega_i$, exhibiting broken power-law distributions.}
    \label{fig:omega}
\end{figure}

Crucially, in Eq.~(\ref{eq:kuramoto}, the driver, or intrinsic frustration term $\omega_i$, is taken as natural frequencies drawn from a broken power-law distribution (to facilitate the non-linear transfer of energy in the ensemble, see Figure~\ref{fig:omega}). $A_0$ is the Kuramoto-coupling strength. As we shall demonstrate, chirality driven by $\omega_i$ plays the role of long-range \textit{inhibitor}, propelling agents away from local clusters, and $Z$ (or $G$) plays the role of short-range \textit{activator}, drawing agents into those clusters. To see how, we point out that the conventional implementation of the Kuramoto model achieves synchronization through a global coupling term between all oscillators: $\sin\left(\phi_i(t) - \phi_j(t)\right)$. To model active chiral matter, we have (as stated in Eq.~\ref{eq:kuramoto}) implemented a locally propagating term, $\sin\left(\phi_i(t) - \Psi(\mathbf{x}_i,t)\right)$, which allows for local inhomogeneities to grow \textit{if} the local mean phase-density $\Psi(\mathbf{x}_i,t)$ is non-zero. Order therefore emerges when inhomogeneities in the spatial distribution of phases randomly align and are amplified through positive feedback, leading to velocity alignment and subsequent spatial clustering of agents.

As we shall demonstrate in the next section, the interplay between inhibition and activation provides the necessary ingredients for a kinetic Turing instability in the ensemble.

Here, we point to Ref.~\cite{luo_turing_2023}, who recently presented a Kuramoto model that operated in a 1D oscillator network, a periodic, discrete chain lattice subject to an explicitly non-linear coupling that involved global two-body attraction and local three-body inhibition. By linearizing the resulting coupled ordinary differential equations around a uniform solution, the authors of Ref.~\cite{luo_turing_2023} demonstrated the presence of Turing instability purely within the coupling topology, with the coexistence of synchronized and incoherent blocks on the chain, or wave-like phase twists, and thereby demonstrating that the the higher-order terms introduced in the coupling term mimics reaction-diffusion dynamics on a network (see also Refs.~\cite{franovic_bumps_2021,zheng_decoding_2025}).

Conversely, our model is a spatial ensemble of active chiral agents, swarming in a periodic bounded 2D box, representing active matter turbulence. The Kuramoto coupling term itself is conventional; attraction and inhibition take place through the curious interplay between \textit{spatial structure} and the agents that \textit{constitute} this structure. We must therefore derive a kinetic extension of the Turing instability into active matter. By so doing, we will bridge the Kuramoto model with plasma physics (Landau damping).

\vspace{-8pt}
\section{Kinetic Turing instability} \label{sec:turing}
\vspace{-8pt}

We shall insert the standard Kuramoto phase-locking dynamic into the kinetic Vlasov-Fokker-Planck equation as the reaction-term, preserving the core Kuramoto physics within a continuum field description. We linearize this equation, which describes long-range interactions in many-body systems, i.e. kinetic plasmas, and the ansatz that we shall use to linearize this equation is identical to the one used to derive Landau damping in plasmas \cite{penrose_electrostatic_1960,strogatz_coupled_1992}. Of some importance, chiral agent frequency, or frustration $\omega_i$, mimics \textit{thermal anisotropy} in kinetic plasmas, while $G$ mimics electric fields. At the same time, the physical transport of phase information involved in the Kuramoto-coupling means that we are explicitly deriving a kinetic instability threshold for Vicsek-swarming and chiral active matter \cite{toner_flocks_1998,wensink_emergent_2012,giomi_defect_2013}, and direct a application to plasma physics remain a future endeavor.

In what follows, we derive a critical coupling strength (that cause agents to cluster rather than chaotically disperse) based on linear stability analysis, following the kinetic approach due to Ref.~\cite{strogatz_coupled_1992} (see also Refs.~\cite{penrose_electrostatic_1960,acebron_kuramoto_2005,okeeffe_oscillators_2017,hong_active_2018}).



\subparagraph{Linear stability analysis}
 
We start by describing the active matter system using the probability density function $F(\mathbf{x}, \theta, \omega, t)$ for oscillators with position $\mathbf{x}$, phase $\theta$, and natural frequency $\omega$ (the latter is drawn from the powerlaw distribution in Figure~\ref{fig:omega}). The evolution is governed by the continuity equation \cite{penrose_electrostatic_1960},
\begin{equation} \label{eq:cont}
    \frac{\partial F}{\partial t} + \nabla \cdot (\mathbf{v} F) + \frac{\partial}{\partial \theta} (v_\theta F) = D \frac{\partial^2 F}{\partial \theta^2},
\end{equation}
where,
\begin{equation} \label{eq:vphase}
    v_\theta = \omega + \frac{A_0}{2i} (Z_{loc} e^{-i\theta} - Z_{loc}^* e^{i\theta}),
\end{equation}
follows from the deterministic part of Eq.~(\ref{eq:kuramoto}), after expanding the $\sin$ term. Eq.~(\ref{eq:vphase}) describes the phase velocity, with the complex field $Z$ mediating the Gaussian kernel interaction (Eq.~\ref{eq:kernel}), now expressed as,
\begin{equation}
    Z_{loc}(\mathbf{x}, t)  = \int_{\mathcal{D}} G(|\mathbf{x} - \mathbf{x}'|) \langle e^{i\theta} \rangle_{\mathbf{x}} d\mathbf{x}',
\end{equation}
$\mathcal{D}$ representing the bounded periodic domain. Here, the term $\langle e^{i\theta} \rangle_{\mathbf{x}}$ is formally given by,
\begin{equation}
    \langle e^{i\theta} \rangle_{\mathbf{x}} = \int \int e^{i\theta} F(\mathbf{x}, \theta, \omega, t) d\theta d\omega
\end{equation}
in the continuum limit rather than the discrete model. Following Ref.~\cite{strogatz_coupled_1992}, we analyze the stability of the incoherent homogeneous state $F_0$,
\begin{equation}
    F_0 = \frac{\rho_0}{2\pi} g(\omega),
\end{equation}
where $\rho_0$ is a uniform spatial density, $g(\omega)$ is the distribution of natural frequencies (the broken power law distributions Figure~\ref{fig:omega}), and where we note that $Z_{loc} = 0$ (no macroscopic order). We then introduce a small perturbation $\delta F$:
\begin{equation}
    F = F_0 + \epsilon \delta F(\mathbf{x}, \theta, \omega, t),
\end{equation}
where $\epsilon$ is a small number. Next, we expand $\delta F$ in Fourier modes for space ($\mathbf{k}$) and phase ($n$):
\begin{equation} \label{eq:exp}
    \delta F = e^{\lambda t} e^{i \mathbf{k} \cdot \mathbf{x}} \sum_{n=-\infty}^{\infty} c_n(\mathbf{k}, \omega) e^{i n \theta},
\end{equation}
where $\lambda$ is the complex growth rate for the perturbation mode, since we assume the perturbation grows as $e^{\lambda t}$. Eq.~(\ref{eq:exp}) is the ansatz, or trial function, to linearize the Vlasov equation \cite{penrose_electrostatic_1960}. The operators of Eq.~(\ref{eq:cont}) are thus simplified,
\begin{equation}
    \frac{\partial}{\partial t} (c_n e^{\lambda t}) = \lambda c_n e^{\lambda t},
\end{equation}
for the time-evolution, and
\begin{equation}
    - D \frac{\partial^2}{\partial \theta^2} (e^{i n \theta}) =  D n^2 e^{i n \theta},
\end{equation}
for the diffusion term, where $D (\partial^2/\partial \theta^2)$ acts on the Fourier mode $e^{i n \theta}$. For the phase advection term we have,
\begin{equation}
    \omega \frac{\partial}{\partial \theta} (c_n e^{i n \theta}) = i n \omega c_n e^{i n \theta},
\end{equation}
and for the spatial gradient, the operator $\nabla$ acting on the spatial wave $e^{i \mathbf{k} \cdot \mathbf{x}}$ pulls down a factor of $i \mathbf{k}$,
\begin{equation} \label{eq:cosnes}
    \mathbf{v} \cdot (i \mathbf{k}) = ik v_0 \cos \theta,
\end{equation}
where we note that \textit{(1)} the gradient runs along the $x$-axis, and \textit{(2)} the factor $\cos \theta = e^{i\theta}/2 + e^{-i\theta}/2$ multiplies the entire sum, splitting it in two.  

Next, Eq.~(\ref{eq:vphase}) enters into the continuity equation (Eq.~\ref{eq:cont}), describing how the Mean Field ($Z$) pushes the incoherent background ($F_0$) into an organized pattern. The expansion,
\begin{equation}
    v_\theta F = (\omega + v_{couple})(F_0 + \delta F),
\end{equation}
eventually yields \cite{strogatz_coupled_1992,acebron_kuramoto_2005}, 
\begin{multline}
    -\frac{\rho_0 g(\omega)}{2\pi} \frac{A_0}{2i} \partial_\theta (Z e^{-i\theta} - Z^* e^{i\theta}) = \\ = \frac{A_0 \rho_0 g(\omega)}{4\pi} (Z e^{-i\theta} + Z^* e^{i\theta}),
\end{multline}
where we note that this driving force acts only on the $n=1$ and $n=-1$ modes, of which $n=1$ is the only physical mode ($c_1$). We project onto $e^{i\theta}$,
\begin{equation}
    (\lambda + D + i\omega + i k v_0 \cos \theta) c_1 = \frac{A_0 \rho_0}{4\pi} g(\omega) Z^*.
\end{equation}
By definition, $Z(\mathbf{k})$ is the convolution of the kernel $\hat{G}(k)$ with the integrated phase density,
\begin{equation}
    Z = \hat{G}(k) \int \int e^{i\theta} \delta F d\theta d\omega = \hat{G}(k) 2\pi \int c_{-1}(\omega) d\omega,
\end{equation}
and assuming $c_{-1} = c_1^*$ (conjugate symmetry of the real perturbation), we solve the $c_1$ equation. Here, we must handle the mentioned angle dependence $\cos \theta$ in the advection term (Eq.~\ref{eq:cosnes}). The effective amplitude $C_1(\omega) = \int c_1 d\theta$ is found by integrating the inverse operator over $\theta$:
\begin{equation} \label{eq:movingosc}
    C_1(\omega) = \frac{A_0 \rho_0 g(\omega)}{4\pi} Z^* \int_{0}^{2\pi} \frac{d\theta}{\lambda + D + i\omega + i k v_0 \cos \theta}.
\end{equation}
Using the identity,
\begin{equation}
    \int_{0}^{2\pi} \frac{d\theta}{A + i B \cos \theta} = \frac{2\pi}{\sqrt{A^2 + B^2}},
\end{equation}
we obtain the exact response function for moving oscillators (Eq.~\ref{eq:movingosc}):
\begin{equation}
    \int_{0}^{2\pi} \dots d\theta = \frac{2\pi}{\sqrt{(\lambda + D + i\omega)^2 + (k v_0)^2}},
\end{equation}
which we subsequently substitute back into the self-consistency condition ($Z^* \propto \int C_1 d\omega$). We then arrive at the exact dispersion relation for the instability onset (see Ref.~\cite{hong_active_2018}),
\begin{equation}
    1 = \frac{A_0 \rho_0}{2} \hat{G}(k) \int_{-\infty}^{\infty} \frac{g(\omega)}{\sqrt{(\lambda + D + i\omega)^2 + k^2 v_0^2}} d\omega.
\end{equation}

Bifurcation occurs when the growth rate $\lambda \to 0$ (marginal stability). The critical coupling strength $A_0$ for a given wavenumber $k$ is then,
\begin{equation} \label{eq:acritexact}
    A_{crit}(k) = \frac{2}{\rho_0 \hat{G}(k) \text{Re} \left[ \chi(k) \right]},
\end{equation}
Where the static susceptibility $\chi(k)$ is given by,
\begin{equation} \label{eq:solvethis}
    \chi(k) = \int_{-\infty}^{\infty} \frac{g(\omega)}{\sqrt{(D + i\omega)^2 + k^2 v_0^2}} d\omega.
\end{equation}
The apt analogue to Landau damping in plasma physics then lies in the breakdown of damping for a particular wavenumber, by the competition between $\hat{G}(k)$ and $\chi(k)$ in Eq.~(\ref{eq:acritexact}).

\begin{figure}
    \centering
    \includegraphics[width=0.5\textwidth]{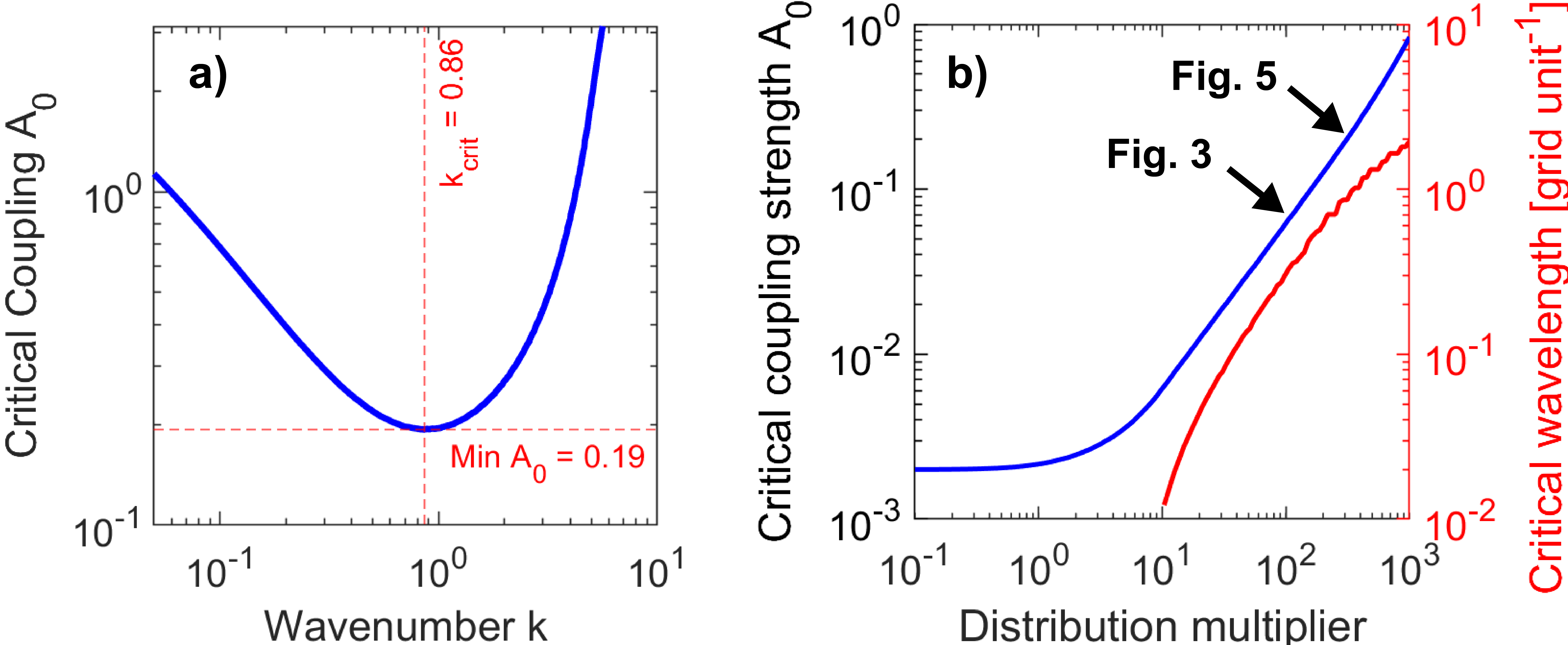}
    \caption{\textbf{Panel a):} Linear stability analysis for the Kuramoto model implemented in the Main Paper, using $\omega$ distribution labeled ``fast mode'' in Figure~1 of the Main Paper. \textbf{Panel b):} Linear stability analysis for all ``distribution multiplier numbers'', that is sliding the oscillation modes from ``slow'' to ``fast'' in Figure 1 in the Main Paper. The configurations used for Figures~\ref{fig:ex} and \ref{fig:stats} are indicated with black arrows.}
    \label{fig:acrit}
    \vspace{-3pt}
\end{figure}

We have implemented a numerical Monte Carlo integration \cite{hamming_numerical_1986} to solve Eqs.~(\ref{eq:acritexact}) and (\ref{eq:solvethis}) using the oscillator frequencies from the distribution of intrinsic agent frustration $\omega_i$ in Figure~\ref{fig:omega}, which we show in Figure~\ref{fig:acrit}a). We observe a minimum for the value $A_0\approx0.2$, for a critical wavenumber of $k_{crit}\approx0.86$. We note a relatively shallow minimum around $A_{crit}$, meaning that a broad range of wavenumbers will be amplified. In Figure~\ref{fig:acrit}b), we show the result of the linear stability analysis applied to all frustration distributions ranging from ``slow'' to ``fast'' in Figure~\ref{fig:omega}, demonstrating that global synchronization ($k_{crit}\to0$ is achieved for distribution multipliers lower than around $10$, at which point global synchronization causes a uniform drift in the agent ensemble. Critical couplings strengths then increase logarithmically with frequency distribution $\omega_i$.

The minimum in Figure~\ref{fig:acrit}a) arises from the interplay between the Gaussian kernel $\hat{G}(k)$ (which suppresses high wavenumbers) and the phase transport or susceptibility $\chi(k)$ (which suppresses low wavenumbers), demonstrating that in our active chiral matter model with frustration, a kinetic Turing instability will select a favorable wavelength and subsequently cause a characteristic structuring of agents. This arrives at a similar result as Ref.~\cite{boccelli_turing_2025}, who utilized the hydrodynamic limit, after likewise identifying the Vlasov-Fokker-Planck equation as a vital to active chiral matter turbulence. The derivation of a critical coupling strength (Eqs.~\ref{eq:acritexact}, \ref{eq:solvethis})  our localized-Kuramoto-active chiral matter model therefore represents a semi-analytical derivation of the onset of active turbulence.


\vspace{-8pt}
\section{Results}
\vspace{-8pt}

We have implemented the model described in Section~\ref{sec:model}, using a coupling strength of $0.2$, a stochastic noise term $\eta_i$ ranging from 0 to 0.15, a constant swim speed of $v_0=0.5$, a Gaussian kernel with interaction radius of 2, and domain boundaries from $-20$ to $+20$ grid units. Agent positions, initial phases, intrinsic frustration $\omega_i$, and noise $\eta_i$ are all drawn random numbers using \textsc{matlab}'s \texttt{philox4x32\_10} stream with seed 45.

To quantify the emergent structuring in the chiral agent ensemble we extract key measurements from the simulation at each timestep. The 1D spatial power spectrum, $P(k)$, of the oscillator point-cloud is calculated using 2D FFT, obtained on the images resulting from binning the oscillator locations onto a 2D $128\times128$-grid. We then azimuthally average the resulting 2D power spectrum. To characterize the geometry of the active turbulence, the log-log power spectrum is fitted with a linear function, and we extract and store its slope, $\alpha$. This is the spectral index, a staple in the characterization turbulent structuring. Its value describes the continuous transfer of power, or energy, from larger to smaller scales, either in shallow (inertial), instability-driven regimes, or in steep dissipative, kinetic regimes \cite{kolmogorov_local_1968,mounirSmallscaleTurbulentStructure1991,spicherPlasmaTurbulenceCoherent2015,ivarsenSteepeningPlasmaDensity2021}.

\begin{figure}
    \centering
    \includegraphics[width=0.495\textwidth]{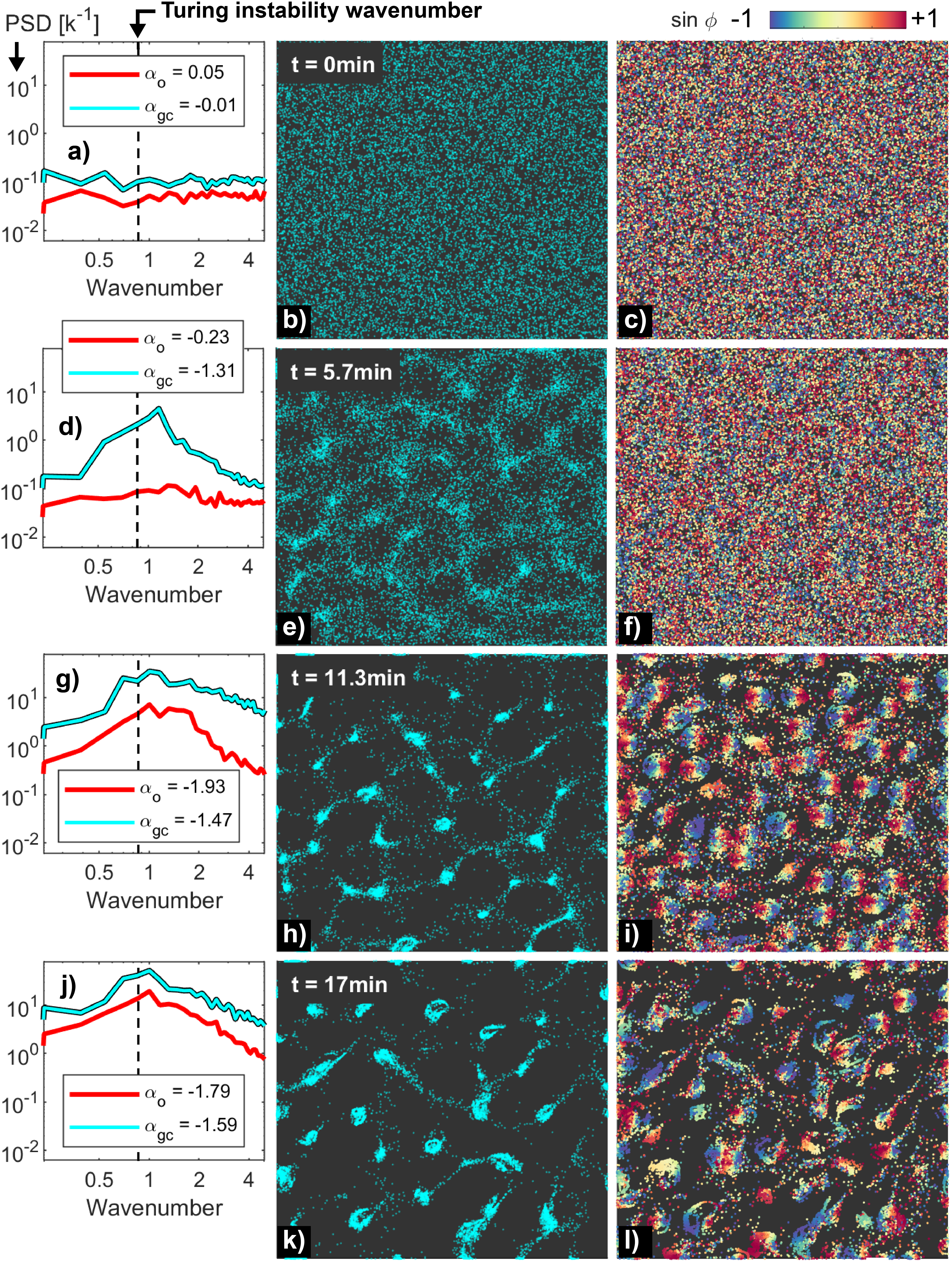}
    \caption{Summary of a simulation with ``fast'' intrinsic frequencies $\omega_i$ and coupling strength $A_0=0.2$. The localized coupling of oscillators is switched on at 2.5~min and kept on. The leftmost column (panels a, d, g, and j) show PSD (cyan for guiding centers and red for active agents), while the middle column (panels b, e, h, and k) show the locations of the oscillators's guiding centers as cyan-colored dots. The rightmost column (panels c, f, i, and l) show the 30,000 oscillators color-coded according to their phase. 
    See Video~S1 in the Supplementary Materials.
    }
    \label{fig:ex}
    \vspace{-3pt}
\end{figure}


Figure~\ref{fig:ex} (and Video S1 in the Supplementary Materials) shows a summary simulation, illustrating the above analysis. The leftmost column shows example powerspectra and their spectral index values. The powerspectra are calculated based on the images on display in the two other columns. The rightmost column shows the positions of the active agents (color-coded by phase $\phi_i$), while the middle column shows the locations of the agents' \textit{guiding centers}. That is, the average location of each agent during a full period of oscillation. Inspecting Eq.~(\ref{eq:kuramoto}) we observe that only the attractive Kuramoto-coupling term can move the guiding centers, and, being averages, they move slowly, moving only after a full cycle in phase (multiples of $2\pi$).  

The wavelength associated with the Turing-instability analysis (Figure~\ref{fig:acrit}a) is indicated with dashed, black line in the first column of Figure~\ref{fig:ex}, and we observe that it roughly coincides with the dominant wavenumber in the structure, providing clear empirical evidence for the analytical derivation of a kinetic Turing instability.

\begin{figure}
    \centering
    \includegraphics[width=0.5\textwidth]{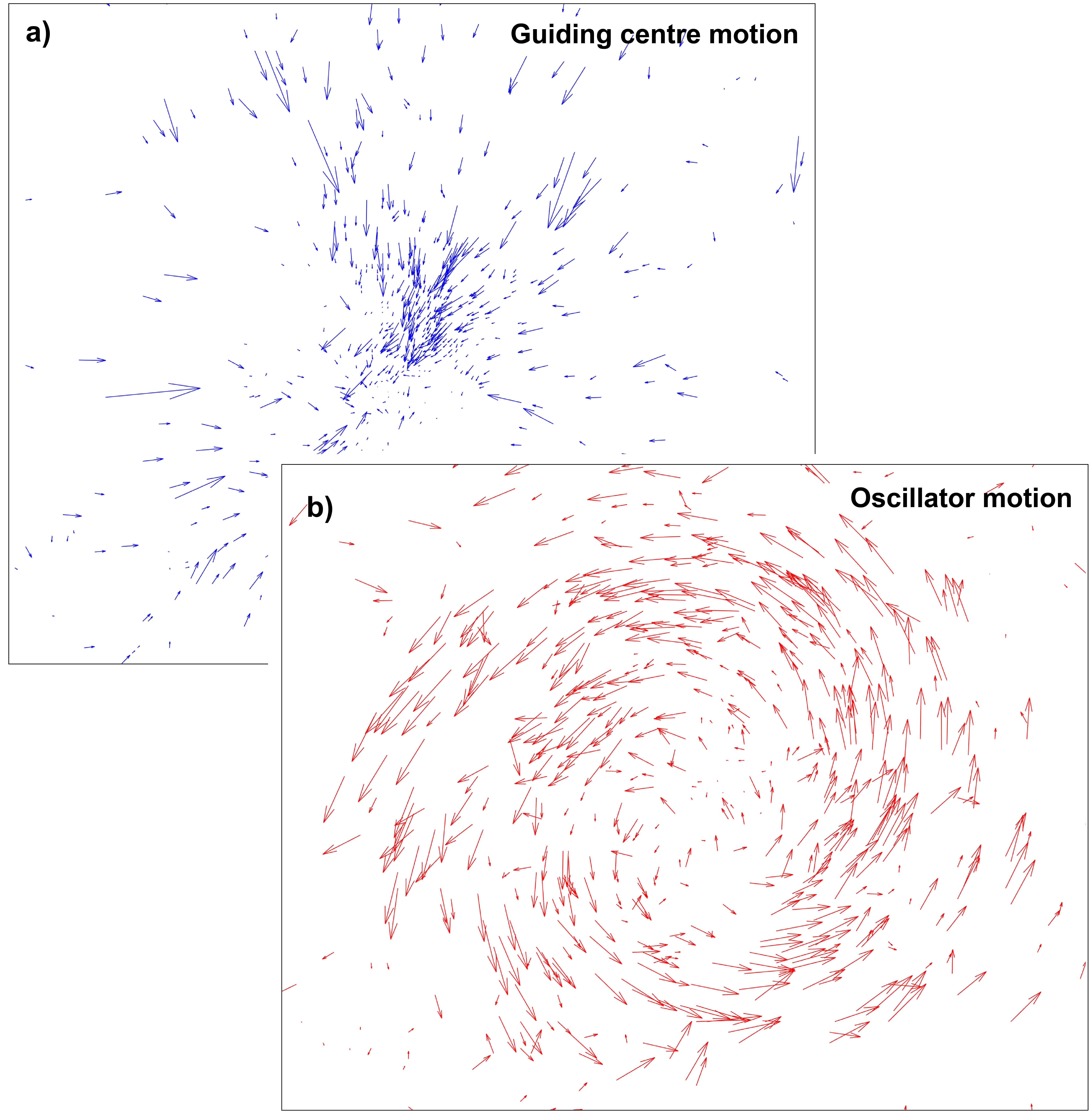}
    \caption{A magnified part of the simulation space, showing the instantaneous velocity of chiral agents that momentarily form a vortex. Panel a) shows the motion of \textit{guiding centers} while panel b) shows the motion of the oscillating agents. The motion arrows in panel a) are scaled up ($\times6.67$) as a visual aid; the motion arrows in panel b) have a fixed magnitude $v_0$, which can be partially averaged out when taking the velocity. See Video~S2 in the Supplementary Materials.
    }
    \label{fig:loop}
    \vspace{-8pt}
\end{figure}

The fast-oscillating modes of Figure~\ref{fig:ex} start in a chaotic, symmetric state (top row). Symmetry breaks with the triggering of the kinetic Turing instability, and the system is subsequently organized in islands with coherent phase evolution, forming local, stable vortices. These vortices must be understood in terms of the quantized loop currents predicted by Ref.~\cite{delabays_multistability_2016}, characterized by the sum of phase differences around closed loop in the network being equal to an integer multiple of $2\pi$. Figure~\ref{fig:loop} 
demonstrates this, showing the instantaneous velocity of some 2000 agents undergoing clustering. 
We observe that tracing a circle around the vortices will cycle through all the phases from 0 to $2\pi$, with a topological winding number of -1 (a counter-clockwise vortex) \cite{delabays_multistability_2016}. We also note that 
\textbf{(1)} the general motion of guiding centers is perpendicular to the motion of the oscillators, and \textit{(2)} the attraction (guiding centers) is considerably slower than the inhibition (agent motility), providing clear justification for the derived Turing instability.

The characteristic spatial organization that we see in Figures~\ref{fig:ex} and \ref{fig:loop} are robust in the simulations of fast-mode oscillators, suggesting that the formation of quantized loop phase currents may be a universal quality in coupled chiral active matter subject to synchronization rules.



\begin{figure*}
    \includegraphics[width=\textwidth]{rev-01.png}
    \caption{The results of a Monte Carlo estimation of the average spectral indices seen in the active chiral matter model. In each simulation, the coupling is initiated at 5\% of the total number of simulation steps and ``turned off" at 67\%. Spectral index is shown with a colorscale (and black contour lines), achieved by systematically varying the base coupling strength $A_0 \in [10^{-2},\pi]$ in logarithmic steps ($y$-axis), using a factor 100 for the natural phase multiplier. Green contour lines indicate the median global order, while a black dash-dotted line indicates the kinetic Turing instability threshold $A_{crit}$. \textbf{Panel a)} shows spectral scaling in \textit{agent} structuring, whereas \textbf{panel b)} shows the spectral index for \textit{guiding center} clustering.
    }
    \label{fig:stats}
    \vspace{-8pt}
\end{figure*}

To substantiate the inferences made thus far, we show in Figure~\ref{fig:stats} aggregates of simulations using a Monte Carlo-based estimate, in which timeseries of the spectral index (and the global order, the mean value of $e^{i\phi_i}$ across $i$) are stacked and averaged across ensembles of simulation runs. 
Figure~\ref{fig:stats}, which uses a distribution in $\omega_i$ slightly ``slower'' than in the foregoing, with a multiplier of $100$, aggregates the results of $24\times24$ simulations. by varying the base coupling strengths $A_0\in[10^{-2},\pi]$. Median spectral index is now shown with a red  (panel a, for the agents) and blue (panel b, for the guiding centers) colorscale, as well as black contour lines of constant spectral index, while global order is similarly shown with green contour lines. 


Figure~\ref{fig:stats} demonstrates that the emergence of structure in the agent ensemble reliably occur for base coupling strengths that exceed the critical value $A_{crit}\approx0.06$ for the kinetic Turing instability, determined after solving Eqs.~(\ref{eq:acritexact}, \ref{eq:solvethis}) for an $\omega_i$ distribution multiplier of $100$. This is heralded by the power spectral density of the spatial distribution of active agents (red) and guiding centers (blue) to settle on power law behaviors.




\vspace{-8pt}
\section{Discussion}
\vspace{-8pt}

We have simulated ensembles of polar chiral active matter agents that are subject to phase synchronization through a Kuramoto-Sakaguchi-like phase interaction, demonstrating that complex structures emerge from a kinetic Turing instability triggered in the relatively simple interactions between locally coupled oscillators. Through the rigorous derivation of a continuum field theory \cite{boccelli_turing_2025} and empirical evidence (Figure~\ref{fig:stats}), we have demonstrated the crucial role played by the kinetic Turing instability within the chiral active matter ensemble. This configuration builds on the noise-multiplicative self-organization framework of Ref.~\cite{marov_self-organization_2013} and is consistent with the theoretical foundation by Ref.~\cite{delabays_multistability_2016} in that stable, quantized loop currents of phase information emerge in the ensemble.

Unlike Ref.~\cite{luo_turing_2023}, where the prerequisites for a Turing instability in a fixed oscillator network were coded into a higher-order Kuramoto interaction, we have demonstrated that a kinetic Turing Turing instability can trigger in the spatial distribution of chiral active matter. Our model relies on a localized conventional Kuramoto interaction, where the attractive force mediated by the kernel  $G$ creates a positive feedback loop that aligns phase-synchronized agents. Opposing this, the agents' swim speed $v_0$ coupled with their intrinsic frequency dispersion $\omega_i$ acts as an effective \textit{diffusion mechanism}, transporting phase information \textit{away} from ordered regions and breaking down order. This creates a competition between local activation and kinetic inhibition, driving the bifurcation observed in the linear stability analysis and triggering the instability.

\begin{figure*}
    \centering
    \includegraphics[width=\textwidth]{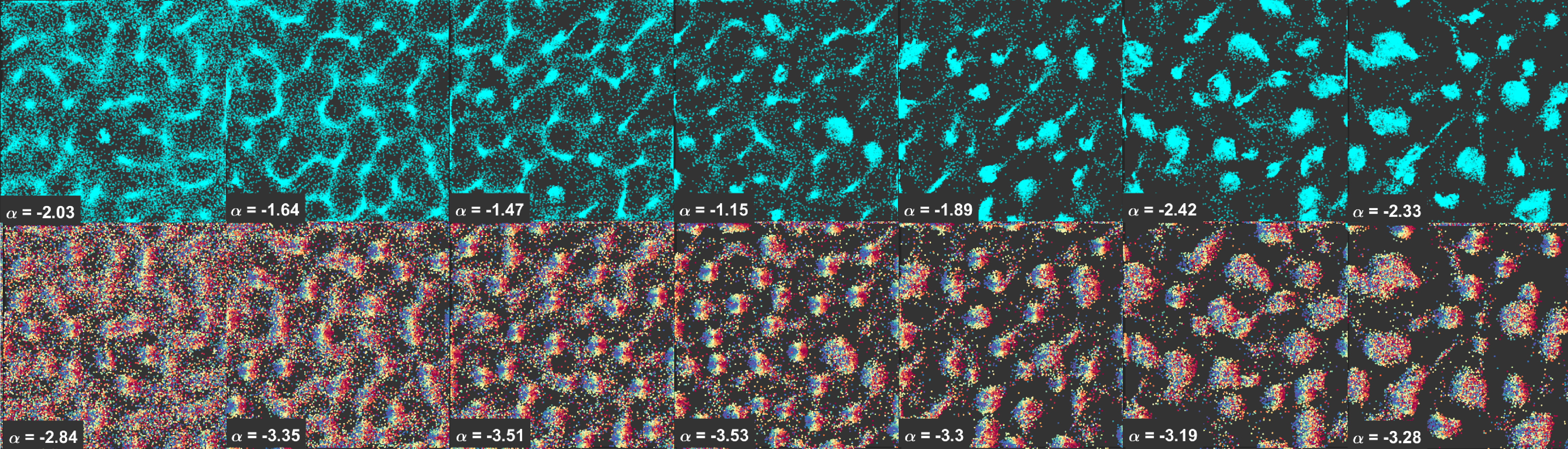}
    \caption{Seven end-state-snapshots of individual fast-mode simulations, using base coupling strengths $A_0$ drawn logarithmically from $[0.07,0.43]$ (from left to right), with otherwise identical initial conditions. The lower row shows oscillator locations with phase color-coded, while the upper row shows guiding center locations in cyan.}
    \label{fig:ensemble}
\end{figure*}

The robust and characteristic structuring caused by the foregoing instability mechanism exhibits stable spectral scaling laws, with spectral index values between $-1.5$ (for the guiding centers) and $-3$ (for the chiral agents). A related paper, Ref.~\cite{ivarsen_onsager_2025-1}, argues that the two regimes reflect the action of a \textit{renormalized fluid element}. Figure~\ref{fig:ensemble} is instructional. It shows a row of single-simulation results using the same data foundation as Figure~\ref{fig:stats}. Going from left to right in Figure~\ref{fig:ensemble}, we increase only the base coupling strength logarithmically through the interval $A_0\in[0.07, 0.43]$, demonstrating the consistent emergence of a filamentary pattern of guiding centers, surrounded by quantized loop currents of phase information.

A crucial ingredient in our model is the heterogenity in $\omega_i$ \cite{marov_self-organization_2013}, the powerlaw distribution in Figure~\ref{fig:omega}. We have confirmed, through rigorous testing, that varying the \textit{shape} (powerlaw) of this distribution does not substantially change the simulation's empirical outcome, and we have verified that the stability analysis is not sensitive to which particular powerlaw we select. Nevertheless, it remains for future efforts to map out the specific mechanism governing the non-linear transfer of energy in the system, replacing the semi-analytical approach in Section~\ref{sec:turing}. In this regard, a related paper, Ref.~\cite{ivarsen_onsager_2025-1}, derives the hydrodynamic limit of our model, and rigorously demonstrates that widening the distribution in $\omega_i$ injects sufficient enstrophy to drive an inertial cascade \cite{marov_self-organization_2013} that eventually culminates in \textit{inviscid Euler turbulence}.

At any rate, the stable spectral index values in Figure~\ref{fig:stats} align with  key observations in active matter turbulence  \cite{kraichnan_inertial-range_1971,borue_spectral_1993,borue_inverse_1994,chertkov_dynamics_2007,reeves_inverse_2013,bratanov_new_2015,giomi_geometry_2015,mecke_emergent_2024}, as well as ionospheric plasma turbulence \cite{kivancSpatialDistributionIonospheric1998,safrankovaPowerSpectralDensity2016,songSpectralCharacteristicsPhase2023,ivarsenDirectEvidenceDissipation2019}, and even kinetic-scale magnetohydrodynamic wave turbulence \cite{boldyrev_spectrum_2012,chenKineticAlfvenWave2013,galtier_entanglement_2015,david_k_perp_2019}, leading us to consider that the emergent spectral scaling laws we have uncovered may be universal, a notion that is bolstered by the governing Vlasov-Fokker-Planck equation \cite{boccelli_turing_2025}, describing long-range interaction across a range of different physical systems.

Our model is, however, constricted to a class of active polar chiral matter, and though the derivation in Section~\ref{sec:turing} explicitly bridges kinetic plasma physics and active matter turbulence modeling, future efforts should explicitly address whether self-organized turbulence can enter into the present reductionist explanations for geophysical and astrophysical turbulent structuring \cite{marov_self-organization_2013}.

\vspace{-8pt}
\section{Conclusion}
\vspace{-8pt}

We have modeled chiral active matter turbulence by coupling self-propelled agents locally with a Kuramoto-like interaction, endowing the agents with an intrinsic and heterogenic frustration, following Ref.~\cite{marov_self-organization_2013}. We have simulated the ensemble, demonstrating that stable spectral scaling laws emerge in the resulting active matter clustering. To explain the results, we derived the spatial pattern formation rules in the active matter, identifying the coupling threshold as a kinetic Turing instability \cite{boccelli_turing_2025}. Following the kinetic theory framework established for active fluids \cite{marchetti_hydrodynamics_2013,doostmohammadi_active_2018,mecke_emergent_2024}, we semi-analytically determined the critical wavenumber selected by the competition between Kuramoto phase-locking and active chiral transport. This provides a rigorous microscopic derivation for the emergent 'chimera-like' patterns often studied in discrete topologies \cite{luo_turing_2023,zheng_decoding_2025}, unifying them under a single continuum instability criterion.

\section*{Acknowledgements}
This work is supported by the European Space Agency’s Living Planet Grant No. 1000012348. The author is grateful to O. Nestande, D. Knudsen, PT. Jayachandran, and K. Douch for stimulating discussions. Google's Gemini 3.0 Pro has been used for mathematical formalism and coding assistance in \textsc{matlab}.


%

\end{document}